\documentclass{article}

\usepackage{arxiv}

\usepackage{fancyhdr}
\usepackage[utf8]{inputenc} 
\usepackage[T1]{fontenc}    
\usepackage{hyperref}       
\usepackage{url}            
\usepackage{booktabs}       
\usepackage{amsfonts}       
\usepackage{nicefrac}       
\usepackage{microtype}      
\usepackage{lipsum}
\usepackage{graphicx}
\graphicspath{ {./images/} }
\usepackage{amsmath}
\usepackage{braket}
\usepackage{graphicx}
\usepackage{subcaption}
\usepackage{amssymb}
\usepackage[numbers]{natbib}

\usepackage{etoolbox}   

\makeatletter
  \patchcmd{\rhead}{\scshape A preprint - \today}{}{}{}

  \patchcmd{\@maketitle}{\center{\today}}{}{}{}
\makeatother

\title{Effectiveness of the DEJMPS purification protocol in noisy entangled photon systems, a Monte Carlo simulation}

\author{
 Vasilis Skarlatos \\
  Department of Informatics\\
  Aristotle University of Thessaloniki\\
  GR-54124, Thessaloniki Greece \\
  \texttt{skarlatov@csd.auth.gr} \\
   \And
 Nikos Konofaos \\
  Department of Informatics\\
  Aristotle University of Thessaloniki\\
  GR-54124, Thessaloniki Greece \\
  \texttt{nkonofao@csd.auth.gr} \\
}
        
\begin{document}
\maketitle
\begin{abstract}
Entanglement purification is a critical enabling technology for quantum communication, allowing high-fidelity entangled pairs to be distilled from noisy resources. We present a comprehensive Monte Carlo study of the DEJMPS purification protocol applied to polarization-entangled photon pairs subject to both amplitude-damping noise ($\gamma$) and dephasing noise ($p$). By sweeping $(\gamma,p)$ over a two-dimensional grid and performing repeated stochastic trials, we map out the average fidelity and average yield surfaces of the purified output, as well as the net gains ($\Delta F$) and losses ($\Delta Y$) relative to the unpurified baseline. Our results show that a single round of DEJMPS purification can boost entanglement fidelity by up to $\approx 0.07$ at the highest-noise corner of the swept grid, while incurring a yield penalty of up to $\approx 0.55$. Fidelity gains grow monotonically with both $\gamma$ and $p$, whereas yields decline more sharply under combined noise. Contour and 3D surface plots of $\Delta F(\gamma,p)$ and $\Delta Y(\gamma,p)$ vividly illustrate the trade-off between quality and quantity of distilled pairs. This two-parameter Monte Carlo characterization provides practical guidance for optimizing purification depth and operating points in real-world photonic networks, and represents, to our knowledge, the first detailed numerical charting of both fidelity and yield improvements across a continuous noise landscape for DEJMPS.
\end{abstract}


\section{Introduction}

The reliable distribution of high‐fidelity entangled photon pairs is a cornerstone of quantum communication and networking. In practice, transmission over realistic channels subjects each photon to both amplitude‐damping noise—characterized by a damping rate $\gamma$—and dephasing noise, with probability $p$; these combined errors degrade entanglement fidelity and undermine protocols such as quantum key distribution and teleportation \cite{Dur1999,Gisin2005}.

Entanglement purification protocols recover quality by processing multiple noisy pairs to distill fewer, higher‐fidelity pairs. The DEJMPS recurrence protocol, based on linear optics and photon counting, remains a leading scheme for photonic platforms \cite{Deutsch1996}. However, purification succeeds probabilistically (“yield”), and the fidelity‐yield trade‐off depends sensitively on both noise parameters and the number of purification rounds.

Recent work has employed Monte Carlo and discrete‐event simulation frameworks to study purification in repeater and network settings \cite{Wallnoefer2024,Coopmans2021}, while surveys of advanced purification techniques have broadened the protocol landscape \cite{Yan2023}. Adaptive protocol‐selection modules have also been proposed to optimize performance under varying input fidelities \cite{Shi2024}. Yet no study has charted the two‐dimensional landscape of DEJMPS fidelity gains and yield penalties across combined amplitude‐damping and dephasing noise in a standalone Monte Carlo setting.

In this work, we introduce a modular Monte Carlo simulation framework that:
\begin{itemize}
  \item Models dual‐channel noise on polarization‐entangled photon pairs,
  \item Applies one or more rounds of the DEJMPS purification protocol,
  \item Sweeps $(\gamma,p)$ over a fine two‐dimensional grid,
  \item Computes average fidelity and yield surfaces for both noisy (unpurified) and purified outputs,
  \item Calculates net fidelity gain $\Delta F$ and net yield change $\Delta Y$ relative to the noisy baseline,
  \item Generates 3D surface plots, contour maps, and cross‐sectional line plots to visualize these metric landscapes.
\end{itemize}

Our code is organized into separate modules for noise modeling, purification routines, Monte Carlo simulation, throughput analysis, and visualization, following a modular design approach inspired by recent work in quantum circuit design and simulation interfaces \cite{KydrosProusalisKonofaos2024}. The resulting two‐parameter maps provide quantitative guidance for choosing purification depth and operating points in near‐term photonic quantum networks.

\section{Review of the DEJMPS Protocol}

The DEJMPS protocol \cite{Deutsch1996} is a recurrence‐type entanglement purification scheme that uses only local operations and classical communication (LOCC) to distill higher‐fidelity Bell pairs from two noisy copies. Each round takes two identical Bell‐diagonal states and probabilistically produces one “better” pair.\\

\begin{figure}[ht]
  \centering
  \includegraphics[width=0.8\linewidth]{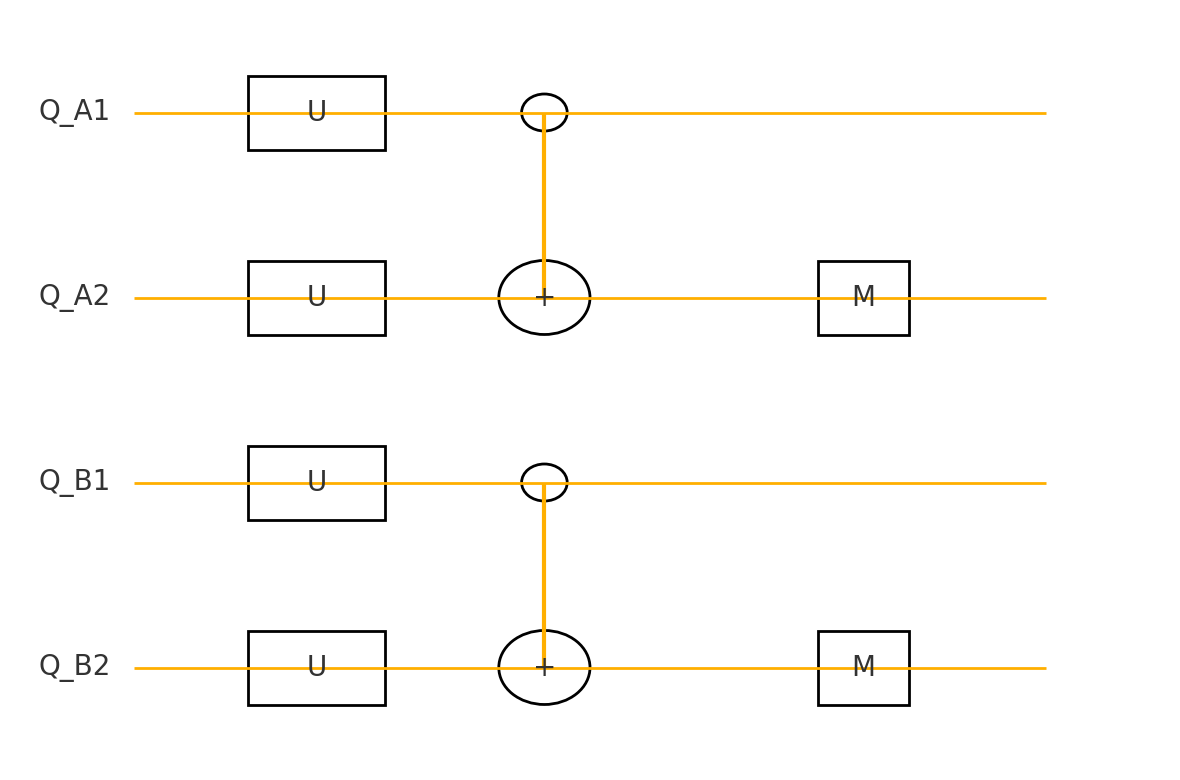}
  \caption{Circuit diagram of one round of the DEJMPS purification protocol. Each party applies a local pre‐rotation $U$ to both qubits, performs a CNOT from the first (control) onto the second (target) qubit, and then measures the target qubits in the computational basis. Success (both measurement outcomes 0) heralds a higher‐fidelity distilled pair.}
  \label{fig:dejmps_circuit}
\end{figure}

The circuit shown in Figure \ref{fig:dejmps_circuit} illustrates a single round of the DEJMPS entanglement purification protocol. Each party begins by applying a bilateral pre‐rotation \(U\) to both qubits, aligning the dominant Bell components for optimal purification. Next, a CNOT gate is executed from the first (control) qubit onto the second (target) qubit, followed by projective measurements of the target qubits in the computational basis. Conditional on both measurement outcomes yielding \(\ket{0}\), the remaining unmeasured pair is retained with an enhanced fidelity. Subsequent rounds can iterate this process to further distill high‐quality entanglement.  

\subsection{Input state}
The starting resource is a Bell‐diagonal two‐qubit state
\begin{equation}
  \rho = \sum_{i=0}^3 \lambda_i \ket{\Phi_i}\bra{\Phi_i}, 
  \qquad
  \sum_{i=0}^3 \lambda_i = 1,
  \label{eq:input}
\end{equation}
where $\{\ket{\Phi_i}\}$ are the four Bell states and the target fidelity is $F = \lambda_0$.

\subsection{Local pre‐rotations}
A bilateral unitary $U\otimes U^*$ permutes the Bell–coefficients so that $\lambda_0$ and $\lambda_3$ become the two largest weights, maximizing the fidelity gain in the next step.

\subsection{CNOT gates \& measurement}
Each party applies a CNOT from their “source” qubit onto their “target” qubit, then measures the targets in the computational basis. The purification succeeds only if both targets are found in $\ket{0}$. The yield of one round is
\begin{equation}
  Y = (\lambda_0 + \lambda_3)^2 + (\lambda_1 + \lambda_2)^2.
  \label{eq:yield}
\end{equation}

\subsection{Post‐selected (output) state}
Conditioned on success, the remaining (source) pair is Bell‐diagonal with updated coefficients
\begin{align}
  \lambda_0' &= \frac{\lambda_0^2 + \lambda_3^2}{N}, &
  \lambda_1' &= \frac{2\,\lambda_1\,\lambda_2}{N}, \nonumber\\
  \lambda_2' &= \frac{\lambda_1^2 + \lambda_2^2}{N}, &
  \lambda_3' &= \frac{2\,\lambda_0\,\lambda_3}{N},
  \label{eq:recurrence}
\end{align}
where
\begin{equation}
  N = (\lambda_0 + \lambda_3)^2 + (\lambda_1 + \lambda_2)^2
  \label{eq:normalization}
\end{equation}
ensures $\sum_i \lambda_i' = 1$.

\subsection{Fidelity update}
Defining input fidelity $F = \lambda_0$ and output fidelity $F' = \lambda_0'$, one round enhances fidelity whenever $F > \tfrac12$.  In general,
\begin{equation}
  F' = \frac{\lambda_0^2 + \lambda_3^2}{(\lambda_0 + \lambda_3)^2 + (\lambda_1 + \lambda_2)^2}.
  \label{eq:fidelity_update}
\end{equation}

\subsection{Iteration \& trade‐off}
By iterating Eqs.~\eqref{eq:recurrence}–\eqref{eq:normalization}, one can approach unit fidelity at the expense of an exponentially decaying yield ($Y^k$ after $k$ rounds). The DEJMPS protocol thus provides a tunable trade‐off between quality (fidelity) and quantity (yield).

\section{Simulation Setup}

The Monte Carlo simulation is implemented and orchestrated by using QuTiP in Python. We model each entangled photon pair by applying amplitude‐damping noise with rate $\gamma$ and dephasing noise with probability $p$ using Kraus operators\cite{Kraus1983States}. For each point on a uniform two‐dimensional grid of $(\gamma,p)$, we perform $N$ independent trials (typically $N=10^4$) in which:

\begin{enumerate}
  \item Two identical noisy Bell pairs are generated.
  \item A single round of DEJMPS purification is applied.
  \item The post‐purification fidelity and success event (yield) are recorded.
\end{enumerate}

We compute the average fidelity $F_{\rm purify}$ and average yield $Y_{\rm purify}$ across trials, as well as the noisy baseline fidelity $F_{\rm noisy}$ (with yield $Y_{\rm noisy}=1$). The net improvements,
\[
\Delta F = F_{\rm purify} - F_{\rm noisy}, 
\qquad
\Delta Y = Y_{\rm purify} - Y_{\rm noisy},
\]
are then visualized using 3D surfaces and contour plots.

\section{Results and Analysis}

\subsection{Average Fidelity and Yield Surfaces}

\begin{figure}[ht]
  \centering
  \begin{subfigure}[b]{0.48\textwidth}
    \centering
    \includegraphics[width=\linewidth]{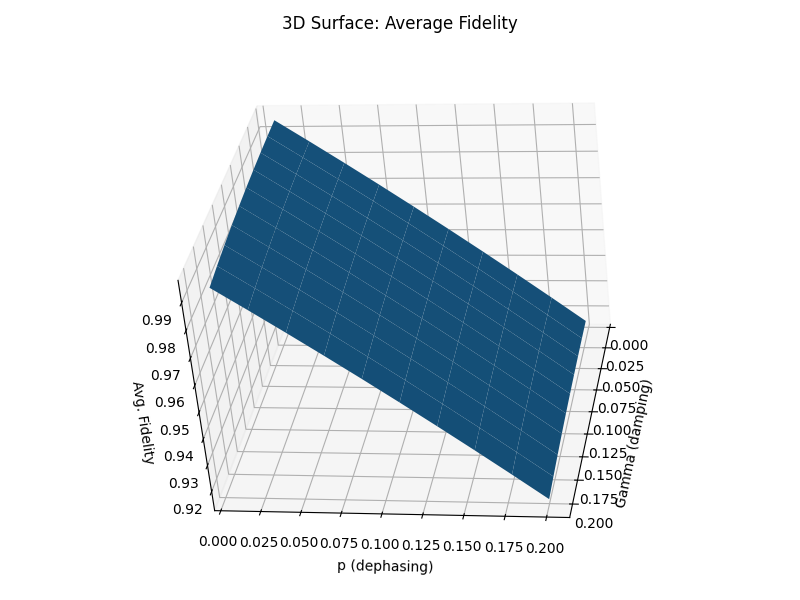}
    \caption{Average purified fidelity $F_{\rm purify}(\gamma,p)$.}
    \label{fig:avg_fidelity_surface}
  \end{subfigure}
  \hfill
  \begin{subfigure}[b]{0.48\textwidth}
    \centering
    \includegraphics[width=\linewidth]{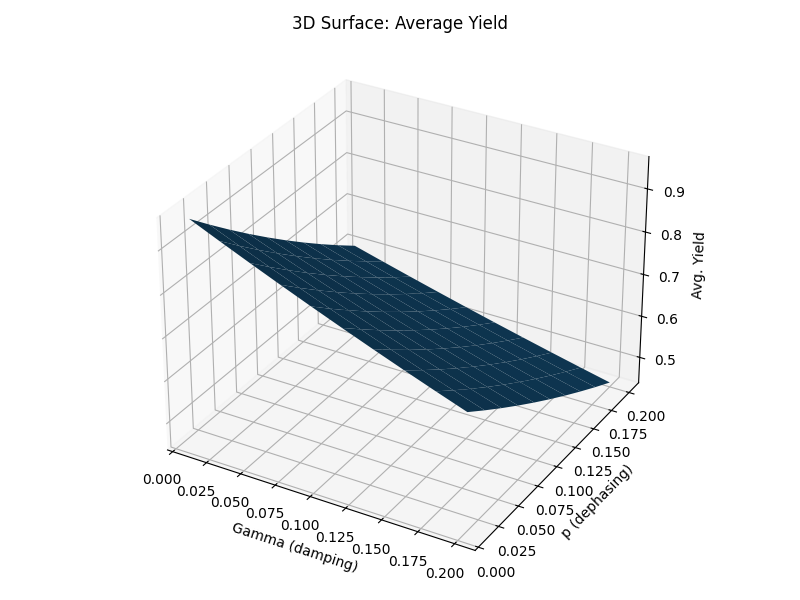}
    \caption{Average purified yield $Y_{\rm purify}(\gamma,p)$.}
    \label{fig:avg_yield_surface}
  \end{subfigure}
  \caption{Average output metrics after one round of DEJMPS purification over the noise parameter grid.}
  \label{fig:avg_surfaces}
\end{figure}

Figure~\ref{fig:avg_surfaces} presents the Monte Carlo–estimated average output fidelity and yield after one round of DEJMPS purification, as functions of amplitude‐damping rate $\gamma$ and dephasing probability $p$. In Figure~\ref{fig:avg_surfaces}(a), the purified fidelity $F_{\rm purify}(\gamma,p)$ remains above $0.9$ across the grid, declining gradually from near unity at $(\gamma,p)\approx(0,0)$ to approximately $0.92$ at $(0.2,0.2)$. This indicates that even under moderate noise, a single purification round sustains high entanglement quality. Figure~\ref{fig:avg_surfaces}(b) shows the corresponding yield $Y_{\rm purify}(\gamma,p)$, which decreases from near unity in the low‐noise region to about $0.45$ at the highest noise. The sharper drop in yield relative to fidelity highlights the probabilistic nature of DEJMPS: purification maintains quality at the expense of throughput under stronger noise.

\subsection{Net Fidelity Gain and Yield Change}

\begin{figure}[ht]
  \centering
  \begin{subfigure}[b]{0.48\textwidth}
    \centering
    \includegraphics[width=\linewidth]{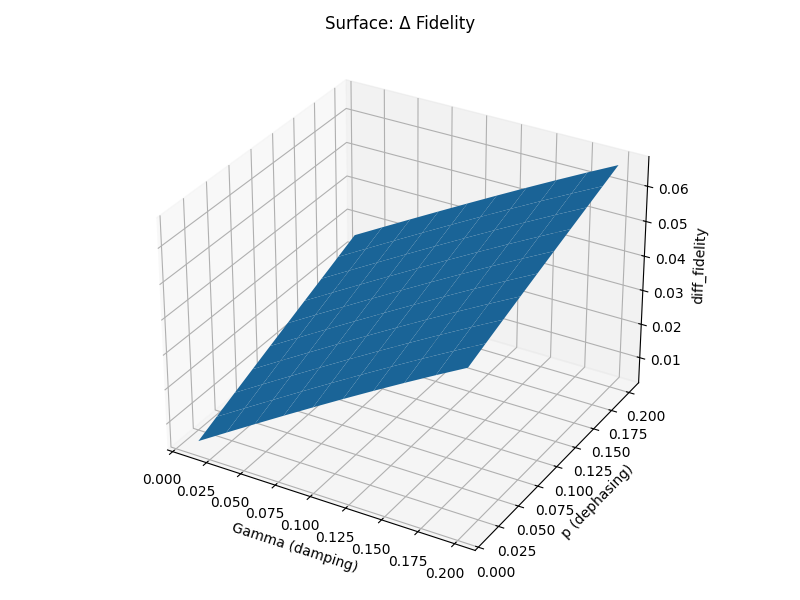}
    \caption{Net fidelity gain $\Delta F(\gamma,p)$.}
    \label{fig:delta_fidelity_surface}
  \end{subfigure}
  \hfill
  \begin{subfigure}[b]{0.48\textwidth}
    \centering
    \includegraphics[width=\linewidth]{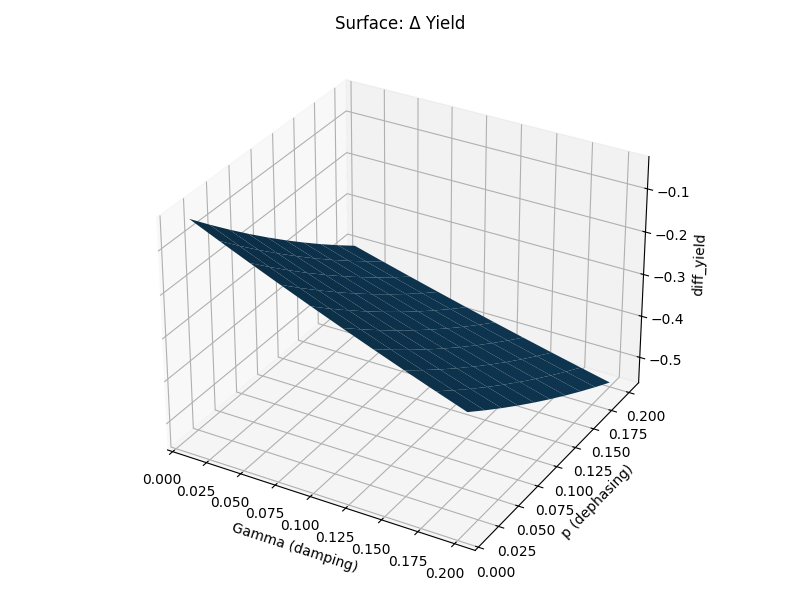}
    \caption{Net yield change $\Delta Y(\gamma,p)$.}
    \label{fig:delta_yield_surface}
  \end{subfigure}
  \caption{Net improvements in fidelity and yield relative to the unpurified baseline.}
  \label{fig:delta_surfaces}
\end{figure}

To quantify the benefit of purification, we define
\[
\Delta F(\gamma,p) = F_{\rm purify}(\gamma,p) - F_{\rm noisy}(\gamma,p),
\qquad
\Delta Y(\gamma,p) = Y_{\rm purify}(\gamma,p) - 1.
\]
Figure~\ref{fig:delta_surfaces}(a) shows $\Delta F(\gamma,p)$ rising monotonically from near zero in the low‐noise regime to a maximum of $\approx 0.07$ at $(\gamma,p)=(0.2,0.2)$, confirming that DEJMPS delivers its greatest fidelity boost exactly where noise is most severe. Figure~\ref{fig:delta_surfaces}(b) plots $\Delta Y(\gamma,p)$, which is always negative and reaches about $-0.55$ at high noise, illustrating the trade‐off: larger fidelity gains come at the cost of reduced success probability.

\subsection{Contour Plots and Cross-Sections}

\begin{figure}[ht]
  \centering
  \begin{subfigure}[b]{0.48\textwidth}
    \centering
    \includegraphics[width=\linewidth]{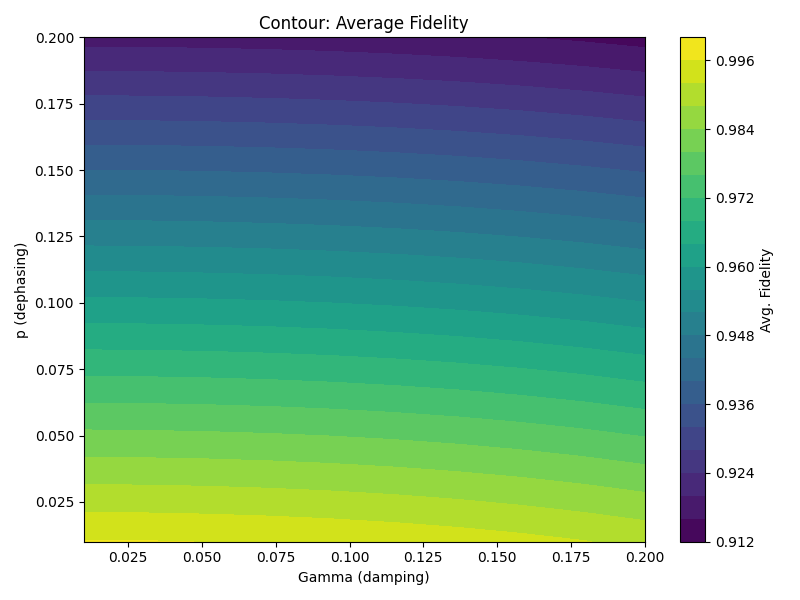}
    \caption{Contour of average purified fidelity $F_{\rm purify}(\gamma,p)$.}
    \label{fig:contour_fidelity}
  \end{subfigure}
  \hfill
  \begin{subfigure}[b]{0.48\textwidth}
    \centering
    \includegraphics[width=\linewidth]{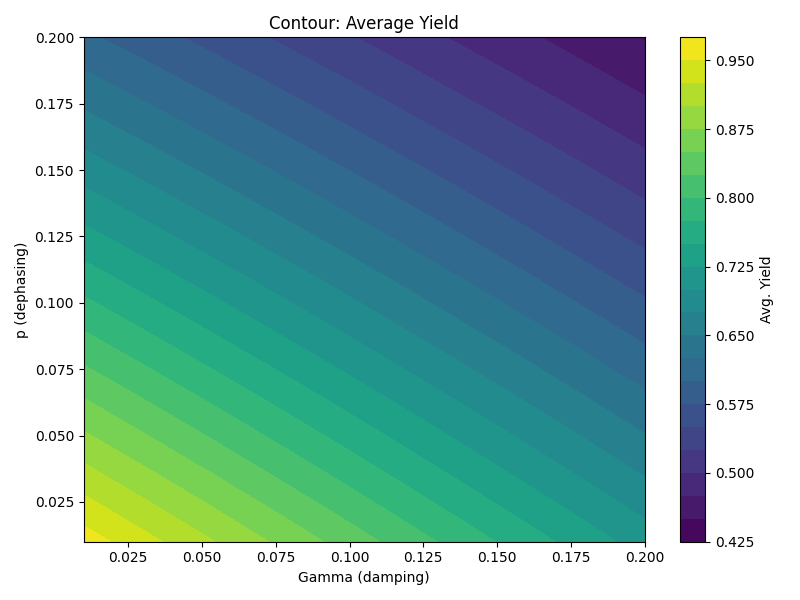}
    \caption{Contour of average purified yield $Y_{\rm purify}(\gamma,p)$.}
    \label{fig:contour_yield}
  \end{subfigure}
  \caption{Contour maps of the average output metrics after one round of DEJMPS purification over the noise parameter grid.}
  \label{fig:contour_surfaces}
\end{figure}

Contour projections in Figure~\ref{fig:contour_fidelity} and Figure~\ref{fig:contour_yield} further delineate operational regimes. Iso‐$\Delta F$ lines show that a target fidelity improvement of $0.03$ can be achieved for $\gamma + p \gtrsim 0.08$, while maintaining a yield $Y_{\rm purify}\ge0.7$ requires $\gamma + p \lesssim 0.07$.

\subsection{Discussion of Trade-offs}

The combined surfaces define a clear trade-off frontier in the $(\gamma,p)$ plane. For low‐noise conditions ($\gamma + p < 0.05$), purification yields minimal fidelity gain with only slight yield loss, suggesting marginal utility. In the intermediate regime ($0.05 \lesssim \gamma + p \lesssim 0.12$), one DEJMPS round achieves fidelity improvements of $0.03$–$0.05$ at yields around $0.8$, striking a balanced enhancement. Under heavy noise ($\gamma + p > 0.12$), fidelity gains exceed $0.06$ but at the expense of sacrificing more than half of the pairs. These quantitative maps empower protocol designers to choose purification depth and operating points that best align with fidelity and throughput requirements in photonic quantum networks.

\section{Discussion and Future Work}

The two‐parameter Monte Carlo maps reveal a clear fidelity–yield trade‐off frontier under joint amplitude‐damping and dephasing noise. Threshold analyses demonstrate that purification yields diminishing returns when input fidelity exceeds a certain level \cite{Dur1999}, while discrete‐event network studies highlight similar probabilistic performance trends in repeater contexts \cite{Wallnoefer2024,Coopmans2021}.

Adaptive multi‐round strategies—where purification depth is chosen dynamically based on real‐time noise estimates—have been proposed in the recent purification literature \cite{Shi2024}, building on the broader framework of quantum error correction \cite{NielsenChuang2000}. Realizing these maps in hardware, on photonic or solid‐state platforms within emerging quantum‐network architectures \cite{Wehner2018}, will be needed to validate our numerical predictions. Coupling with large‐scale network simulators enables end‐to‐end analysis, and extending these methods to hyperentangled and higher‐dimensional protocols may unlock new regimes of entanglement distribution efficiency.

\medskip

\noindent\textbf{Future work} will investigate:
\begin{itemize}
  \item Dynamic selection of purification rounds via feedback control.
  \item Experimental validation on photonic and semiconductor platforms.
  \item Integration with correlated‐noise models for more realistic channel simulations.
  \item Extension to hyperentanglement and multipartite purification schemes.
\end{itemize}

\bibliographystyle{unsrtnat}  
\bibliography{references}  

\section{Acknowledgments and Data Availability}

The authors declare no competing interests.\\

We acknowledge and thank Zixuan Lu for their work on the \LaTeX \space template on which this article was written.\\

The code used for the simulations and derivations of the data used for the graphs shown in this paper can be found at \url{https://github.com/BillSkarlatos/Purification_Simulation} and be accessed and reproduced freely.

\end{document}